\documentclass[11pt]{article}

\title{Quantum particles and an effective spacetime geometry}

\author{Yuri Bonder}

\date{Instituto de Ciencias Nucleares\\ Universidad Nacional Aut\'onoma de M\'exico\\
A. Postal 70-543, M\'exico D.F. 04510, M\'exico\\ yuri.bonder@nucleares.unam.mx}

\begin{document}
\maketitle

\section*{Abstract}
Spacetime geometry is supposed to be measured by identifying the trajectories of free test particles with geodesics. In practice, this cannot be done because, being described by Quantum Mechanics, particles do not follow trajectories. As a first step to study how it is possible to read spacetime geometry with quantum particles, we model these particles with classical extended objects. We propose to represent such extended objects by its covariant center of mass, which generically does not follow a geodesic of the background metric. We present a scheme that allows to extract some of components of an ``effective'' connection, namely, the connection that would be obtained if the locus of the center of mass is regarded as a geodesic. We discuss some issues that arise when trying to obtain all the components of the effective connection and its possible implications.

\section{Introduction}

Point-like particles play a crucial role when interpreting spacetime as a geometrical entity. For example, according to General Relativity, geodesics are the paths that are followed by free test point-like particles. On the other hand, real particles satisfy the principles of Quantum Mechanics which forbid to localize a particle on a path. This apparent incompatibility between General Relativity and Quantum Mechanics has not been fully explored and it could reveal some clues on how to formulate a fully consistent theory of Quantum Gravity. Furthermore, it is interesting to investigate if spacetime, as a pseudo-Riemannian manifold, can be rigorously defined avoiding the idealization of classical point particles.

Tackling the problem described above is technically complicated. Thus, as a first step to gain conceptual insight, we have replaced the wave-function of a test particle by an extended classical object which is characterized by its covariant center of mass. The strategy is to simplify the question and we concentrate in exploring how the extended nature of realistic test objects modify the way we measure geometry, assuming that a (classical) underlaying spacetime geometry exist. As is well known \cite{Papapetrou}, even if an extended object is free, the locus of its center of mass is generically not a geodesic. Our goal is to analyze if it is possible to extract an effective (\textit{i.e.}, object dependent) spacetime geometry when representing extended objects by its center of mass. In this manuscript we focus in giving a definition of an effective connection, namely, a connection that, when used in the geodesic equation instead of the usual Levi-Civita connection, has as solutions the center of mass world-lines.

The manuscript is organized as follows: We first introduce the covariant definition of the center of mass. Then we derive the effective connection. We finish the paper with some concluding remarks. This work is written following the notation and conventions of Ref. \cite{Wald}.

\section{Covariant center of mass in curved spacetimes}

The center of mass in Special Relativity was studied long ago \cite{Pryce}. The main difficulties in taking this definition to curved spacetimes is that in this case the position of each ``piece'' of the extended object cannot by simply given by its coordinates and that, in order to sum vectors, they must be at the same tangent space. Regardless of these difficulties, Dixon \cite{Dixon} was able to give a covariant\footnote{In this work covariance is meant to be not only independence of the coordinates but also from an observer. The notion of a \textit{centroid} could be thought as the center of mass associated with a particular observer and it is still coordinate independent.} prescription of  center of mass in curved spacetimes, which is essentially the one we use in this paper.

The precise hypothesis guaranteeing that the center of mass in curved spacetimes is well and uniquely defined where found by Beiglb\"ock \cite{Beiglbock}. Loosely speaking, one has to assume that the radius of curvature in the region $\mathcal{O}$ where the extended object is, is big in comparison with the size of the extended object. More precisely, the region $\mathcal{O}$ has to be a normal convex hull, \textit{i.e.}, any pair of points in $\mathcal{O}$ must be connected by a unique geodesic that is entirely contained in $\mathcal{O}$. In addition, we assume that we know spacetime metric $g_{ab}$. For simplicity, we consider the extended objects to be a collection of $N$ point-like particles; the generalization to a continuous distribution of matter is straightforward.

To calculate the center of mass we take an arbitrary point $x_0\in \mathcal{O}$ and an arbitrary $4$-velocity $U_0^a \in V_{x_0}$. The simultaneity hyper-surface with respect to $U_0^a$, denoted by $\Gamma(x_0,U_0)$, is given by all the geodesics that pass through $x_0$ and whose tangents at this point are orthogonal to $U_0^a$. Denote by $y_{(i)}(x_0,U_0)$ to the point where the $i$-th particle world-line intersects $\Gamma(x_0,U_0)$. By assumption, there is only one geodesic connecting $y_{(i)}(x_0,U_0)$ and $x_0$. This allows to unequivocally find the vectors $\xi_{(i)}^a(x_0,U_0)\in V_{x_0}$ such that
\begin{eqnarray}
\label{cond 1} g_{ab}(x_0)U^a_0 \xi_{(i)}^b(x_0,U_0)&=&0,\\ 
\label{cond 2}\exp_{x_0}\left[\xi^a_{(i)}(x_0,U_0)\right]&=&y_{(i)}(x_0,U_0),
\end{eqnarray}
where $\exp_{x}$ is the exponential map which assigns to a vector $v^a \in V_x$ the point one gets after following an affine distance $1$ the geodesic that emanates from $x$ with tangent $v^a$. Note that $\xi_{(i)}^a(x_0,U_0)\in V_{x_0}$ may be regarded as the vector position of the $i$-th particle with respect to $x_0$ and at the instant determined by $U_0^a$.

We can take the momentum of the $i$-th particle from $y_{(i)}(x_0,U_0)$ to $x_0$ by parallel transport along the unique geodesic joining these points, the result of this operation is denoted by $\tilde{p}^a_{(i)}(x_0,U_0)$. The total momentum at $x_0$ and with respect to $U_0^a$ is defined as
\begin{equation}
\tilde{P}^a(x_0,U_0)= \sum_{(i)=1}^N\tilde{p}^a_{(i)}(x_0,U_0).
\end{equation}
This last definition can be used to get a preferential $4$-velocity at $x_0$, $U^a(x_0)\in V_{x_0}$, which can be though as the $4$-velocity of an observer that sees the extended object at rest. The idea is to look for the $U^a(x_0)$ such that the total momenta at $x_0$ with respect to $U^a(x_0)$ is parallel to it, namely,
\begin{equation}
U^a(x_0)\propto \tilde{P}^a(x_0,U(x_0)).
\end{equation}
It has been shown \cite{Beiglbock} that under the work hypothesis, $U^a(x_0)$ exists and it is unique. We define \textit{the} momentum of the $i$-th particle at $x_0$ and the total momentum (independent of an arbitrary $4$-velocity) as
\begin{equation}
p_{(i)}^a(x_0)=\tilde{p}^a_{(i)}(x_0,U(x_0)),\qquad
P^a(x_0)=\tilde{P}^a(x_0,U(x_0)).
\end{equation}
The preferential $4$-velocity allows to define an energy for each point-particle:
\begin{equation} \label{def E}
E_{(i)}(x_0)=-g_{ab}(x_0)p_{(i)}^a(x_0)U^b(x_0).
\end{equation}
Moreover, given that $\xi_{(i)}^a(x_0)=\xi_{(i)}^a(x_0,U(x_0))$ is the vector position of the $i$-th particle with respect to $x_0$, the ``vectorial'' center of mass \textit{with respect to $x_0$} can be defined as
\begin{equation}
X^a(x_0)= \frac{\sum_{(i)=1}^N \xi_{(i)}^a(x_0) E_{(i)}(x_0) }{\sum_{(j)=1}^N E_{(j)}(x_0) }.
\end{equation}
Although not necessary for the center of mass calculation, it is convenient to define, at this point, the total angular momentum of the extended object with respect to $x_0$, which is given by
\begin{equation} \label{def J}
J^{ab}(x_0)=2 \sum_{(i)=1}^N \xi_{(i)}^{[a}(x_0) p^{b]}_{(i)}(x_0).
\end{equation}

Clearly the center of mass needs to be a spacetime point and to get a point out of $X^a(x_0)$ we use the exponential map
\begin{equation}
\mathcal{S}(x_0)= \exp_{x_0}[X^a(x_0)].
\end{equation}
In addition, the center of mass has to be independent of the arbitrarily chosen point $x_0$. This is achieved by extending\footnote{Note that, when extending the $\mathcal{S}$ map over $\mathcal{O}$, all the vectors/tensors we define above become vector/tensor fields in $\mathcal{O}$.} $\mathcal{S}$ to every $x\in \mathcal{O}$ to get a map $\mathcal{S}:\mathcal{O}\rightarrow\mathcal{O}$. \textit{The} center of mass world-line is defined as the set of fixed points of $\mathcal{S}$, namely, the curve $Z\in\mathcal{O}$ such that
\begin{equation}\label{fixed point cond}
Z=\mathcal{S}(Z).  
\end{equation}
It is important to mention that $Z$ exists in a unique way and it is a differentiable time-like curve \cite{Beiglbock}. In addition, it is possible to check that this definition has the correct non-relativistic limit.

\section{Effective connection} 

In this section we study if, for a given extended object, there is an ``effective'' connection such that the solutions of the geodesic equation obtained with it coincide with the center of mass world-line. In order to do so, we use the fact that equation (\ref{fixed point cond}) implies
\begin{equation} \label{fixed point 2}
\sum_{(i)=1}^N \xi_{(i)}^a(Z) E_{(i)}(Z) =0.
\end{equation}
Let $W^a=g_{bc}P^b J^{ca}$. The important point is that, by combining equations (\ref{cond 1}) and (\ref{fixed point 2}), it can be proven that\footnote{This is the expression that used Dixon to give the covariant definition of the center of mass in curved spacetimes and it has motivated alternative definitions \cite{Madore}.}
\begin{equation} \label{W=0}
W^a(Z)=0.
\end{equation} 
This is the property that allows us to define the effective connection

As $W^a$ vanishes along $Z$, so do its (covariant) derivatives. Denoting by $\dot{Z}^a$ the tangent of the center of mass, the ``second'' derivative of $W^a$ in the direction $\dot{Z}^a$ satisfies
\begin{eqnarray}
0&=&\dot{Z}^c \nabla_c(\dot{Z}^b \nabla_b W^a)\vert_Z\nonumber\\
&=&\dot{Z}^c \nabla_c\dot{Z}^b\vert_Z (\nabla_b W^a)\vert_Z+\dot{Z}^b \dot{Z}^c \nabla_c\nabla_b W^a\vert_Z\nonumber\\
&=&\ddot{Z}^b \partial_b W^a\vert_Z+\dot{Z}^b \dot{Z}^c \left(\partial_c\partial_b W^a+2 \Gamma^a_{bd} \partial_c W^d\right)_Z,\label{geom efect}
\end{eqnarray}
where in the last step we use equation (\ref{W=0}) and we define $\ddot{Z}^a=\dot{Z}^b\partial_b \dot{Z}^a$. The effective connection we are looking for (associated with the coordinates used and parametrization of the center of mass world-line) is defined as
\begin{equation}\label{def Gamma tilde}
0=\ddot{Z}^\mu + \dot{Z}^\rho\dot{Z}^\sigma \widetilde{\Gamma}_{\rho\sigma}^\mu(Z).
\end{equation}
Provided that $\partial_\mu W^\nu\vert_Z$ is invertible and writing its inverse as $M_\mu^\nu$, equation (\ref{geom efect}) takes the form
\begin{equation}
0=\ddot{Z}^\mu +\dot{Z}^\rho \dot{Z}^\sigma \left(\partial_\rho\partial_\sigma W^\nu+2 \Gamma^\nu_{\rho \kappa} \partial_\sigma W^\kappa\right)_Z M_\nu^\mu.
\end{equation}
By comparing these last equation with equation (\ref{def Gamma tilde}), it is tempting to conclude the effective connection at $Z$ is given by
\begin{equation}\label{gamma tilde}
\tilde{\Gamma}_{\rho\sigma}^\mu(Z)=\left(\partial_\rho\partial_\sigma W^\nu+2 \Gamma^\nu_{\kappa(\rho} \partial_{\sigma)} W^\kappa\right)_Z M_\nu^\mu.
\end{equation}
It is important to note that $\tilde{\Gamma}_{\rho\sigma}^\mu(Z)$ depends on the linear and angular momentum of the extended object through $W^a$.

A closer inspection to this derivation allows us to note that we only obtain the components of $\tilde{\Gamma}_{\rho\sigma}^\mu(Z)$ along the center of mass world-line. In other words, we can add to equation (\ref{gamma tilde}) any tensor $\gamma_{\rho\sigma}^\mu$ such that $\gamma_{\rho\sigma}^\mu\dot{Z}^\rho \dot{Z}^\sigma=0$. This is not surprising if we recall that when probing a spacetime point with one point-like particle, we only get the components of the ``true'' connection along the particle's tangent. However, if we manage to test this point with enough point-like particles it is, in principle, possible to extract all the components of this connection. The additional complication faced when considering extended objects is that the effective connection depends on the extended object and, in order to probe different directions, we need to use different extended objects (having, at least, different momenta). \textbf{Thus, each extended object measures some components of a different effective connection and not different components of the same effective connection}. This results seems to be saying that the effective connection, as an object having information about \textit{all} possible trajectories of the center of mass for a given extended object cannot be operationally defined, but only its components along the center of mass world-line.

\section{Conclusions}

In this work we analyze if some geometrical features of spacetime can be consistently defined when the quantum nature of the particles is taken into the account. We model quantum probing particles by free extended classical objects and we assign its covariant center of mass as the point that represents them. Moreover, we also assume that the background (classical) spacetime metric is known. Motivated by the fact that the center of mass generically does not follow a geodesic of the background metric, we present a derivation of an effective connection, \textit{i.e.}, an object entering in a geodesic equation whose solutions coincide with the center of mass of a given extended object. In doing so we find that only the components of the effective connection along the center of mass curve can be obtained since several (different) extended objects are needed to probe all spacetime directions. 

This problem could be solved by characterizing when extended objects in a curved spacetime can be considered as equal. One possibility is to restrict to spacetime regions with a flat region in its past where copies of an extended object could be prepared. Still, each copy will deform in a different way before getting to the point we want to probe. An other interesting idea is to send extended objects that, in the point we want to probe, are described by the same covariant quantities (\textit{e.g.}, total mass, proper size, etc.) Further investigations are needed to explore these ideas.

Even if we find a method to obtain all the components of the effective connection, we would need to verify if it is possible to reconstruct from it an effective metric. Certainly, this seems to be a nontrivial problem. One possible solution in this direction would be to extend the effective connection to a neighborhood in such a way that it would be possible to calculate an effective Riemann tensor. Then, an effective metric could be defined by using normal Riemann coordinates associated with the point where the effective Riemann tensor is known.

The main lesson from this study is that to reconcile the geometrical aspects of General Relativity with the fact that quantum particles do not follow trajectories is a formidable task and, in trying to do so, we have found several results suggesting that it may be impossible to resolve this issue. If this is the case, it is possible that this incompatibility would manifest empirically before the Planck scale. In addition, it is expected that this same problem will arise in any Quantum Gravity theory when trying to recover the classical limit, placing a new challenge to these theories.

\section*{Acknowledgments}
This work was partially supported by the research grants CONACyT 101712 and 103486 and PAPIIT-UNAM IN107412.

\end{document}